\documentclass [12pt,reqno]{article}%
\usepackage{amsmath}
\usepackage{amsfonts}
\usepackage{amssymb}
\usepackage{graphicx}%
\newcommand{\bpartial}{\mathop{\partial\kern -4pt\raisebox{.8pt}{$|$}}}
\newcommand{\bra}{\mathopen{[\kern-1.6pt[}}
\newcommand{\ket}{\mathclose{]\kern-1.5pt]}}
\newcommand{\bbra}{\mathopen{[\kern-2.2pt[\kern-2.3pt[}}
\newcommand{\bket}{\mathclose{]\kern-2.1pt]\kern-2.3pt]}}

\makeindex
\begin{document}

\title {\large{\bf Integrable and superintegrable Hamiltonian systems with four dimensional real Lie algebras as  symmetry of the systems}}

\vspace{3mm}

\author { \small{ \bf J. Abedi-Fardad$^1$ }\hspace{-2mm}{ \footnote{ e-mail:j.abedifardad@bonabu.ac.ir}}, \small{ \bf A.
Rezaei-Aghdam$^2$
}\hspace{-2mm}{\footnote{ e-mail: rezaei-a@azaruniv.edu}},\small{ \bf  Gh. Haghighatdoost$^3$}\hspace{-2mm}{ \footnote{ e-mail:gorbanali@azaruniv.edu}}\\
{\small{$^{1,3}$\em Department of Mathematics, Bonab
University , Tabriz, Iran}}\\
{\small{$^{2}$\em Department of Physics, Azarbaijan Shahid Madani
University, 53714-161, Tabriz, Iran}}\\
 {\small{$^{3}$\em Department
of Mathematics,Azarbaijan Shahid Madani University, Tabriz, Iran}} }
 \maketitle

\begin{abstract}
We construct integrable and superintegrable Hamiltonian systems
using the realizations of four dimensional real Lie algebras  as a
symmetry of the system with the  phase space $\mathbb {R}^{4}$ and
$\mathbb {R}^{6}$ . Furthermore, we   construct some  integrable and
superintegrable Hamiltonian systems for which the symmetry  Lie
group is also the phase space of the system.
\end{abstract}
{\bf keywords}:Integrable Hamiltonian systems, Superintegrable
Hamiltonian systems, Lie algebra.
\section{\bf Introduction}
A Hamiltonian system with $N$ degrees of freedom is integrable from the Liouville sense if it has N invariants in involution (globally defined and functionally independent);\cite{FM} and is superintegrable if it has additional independent invariants up to $2N-1$ . Superintegrablility forces analytic and algebraic solvability. The modern theory of superintegrability was pioneered by Smorodinsky, Winternitz and collaborators\cite{pt} (see for recent review \cite{msp}). \\
In this work, we construct new integrable and superintegrable
Hamiltonian systems by using the realizations of four dimensional
real Lie algebras \cite{rop} as a  symmetry of the system with the
phase space $\mathbb {R}^{4}$ and $\mathbb {R}^{6}$. Furthermore by
use of these realizations we  construct integrable and
superintegrable Hamiltonian systems on symmetry  Lie groups as phase
space. Note that previously in \cite{aak} some integrable
Hamiltonian systems were constructed  on low dimensional real Lie
algebra with their coalgebra as phase space.  In that work, the
invariants of the systems were not specified as a function of phase
space variable.

\section{\bf Integrable systems with phase space $\mathbb {R}^{4}$ and $\mathbb {R}^{6}$ }

Here, we use the  classification of four dimensional real Lie
algebra ($A_4$) which has been presented in   \cite{jp}, and
construct integrable Hamiltonian systems with the phase space
$\mathbb {R}^{4}$ or $\mathbb {R}^{6}$ such that the Casimir
invariants of these Lie algebras are Hamiltonians of the systems.
For this proposes, we consider the function $Q_i$ ~~(
$i=1,...,$dimension phase space) of the phase space ($\mathbb
{R}^{4}$ or $\mathbb {R}^{6}$) variables $(x_a, p_a )$ such that
they satisfy the following Poisson brackets:
\begin{equation}
\lbrace Q_i,Q_j \rbrace =f_{ij}^k Q_k   ~~,
\end{equation}
where $f_{ij}^k$ are  the structure constants of the symmetry Lie
algebra. Then one can consider the Casimir of the Lie algebra as
Hamiltonian of the system where the dynamical observable $Q_i^{~,}s$
replaced with the generators of the Lie algebra in the Casimir. For
obtaining the functions of $Q_i$ we use the differential realization
of the Lie algebras $A_4$ \cite{rop} such that in these realizations
we replace the $\partial_{x_i}$ with the  momentum $p_{i}$.

Now let us consider an example; for Lie algebra $ A_{4,1}$ according
to \cite{rop} we have the following commutators and realization on
$\mathbb{R}^6$:

 \begin{equation}
   X_1=\partial_1  ~,\quad X_2=\partial_2 ~ , \quad X_3=\partial_3 ~,\quad X_4=x_2 \partial_1+x_3\partial_2 ~ ,
 \end{equation}

 \begin{equation}
   [X_2,X_4]=X_1~,~~[X_3,X_4]=X_2 ~,
 \end{equation}\\
 where $x_i $ are coordinates of  $\mathbb {R}^{6}$ and $\partial_i\equiv\frac{\partial}{\partial{x_i}}$.\\
Then, we construct the following $Q_i^{~,}s$,  $i=1,2,3,4$ as a
function of $(x_1,x_2,x_3,p_1,p_2,p_3) $ variables of $\mathbb
{R}^{6}$  phase space from the above realization such that they have
the following forms and  Poisson brackets:
\begin{equation}
Q_1=-p_1~,\quad Q_2=-p_2~,~\quad Q_3=-p_3~,\quad Q_4=-x_2 p_1- x_3
p_2~,
\end{equation}
\begin{equation}
\lbrace Q_i,Q_j \rbrace =f_{ij}^k Q_k~,
\end{equation}
where $f_{ij}^{~k} $ is  the structure constants \cite{rop} of the
Lie algebra  $ A_{4,1}$. Now, with the above form for
 $Q_i^{~,}s$ the  Casimir of Lie algebra $A_{4,1}$ \cite{jp} as a Hamiltonian of the system has the following form:
\begin{equation}\label{a1}
{ H = Q_2^2-2Q_1Q_3=p_2^2-2 p_1 p_2}~.
\end{equation}

 In this way, we construct a \textit{superintegrable} system with Hamiltonian (\ref{a1}) and invariants $(H,Q_1,Q_2,Q_3)$ on the phase space $\mathbb {R}^{6}$. The results for other four dimensional real Lie algebras are summarized in the table 1 and 2. In table 1 we summarized the integrable and superintegrable systems with phase space $\mathbb {R}^{4}$ and their symmetry Lie algebras. The result of above work with phase space $\mathbb {R}^{6}$ are summarized in table 2.\\
\newpage

{
\qquad Table 1:{\tiny{} Integrable and superintegrable systems with the phase space $\mathbb {R}^{4}$. }}\\
{\tiny
\begin{tabular}{|p{2.78cm}|p{3.5cm}|p{3.8cm}|p{1.65cm}|}
\hline {symmetry Lie algebra} &  & &\\
{(nonzero commutation}   &\hspace{1.5cm} $Q_i$&\hspace{1.75cm}  $H$& invariants  \\
\qquad relations)&  & &\\\hline
 $A_{4,1}  $ & $ Q_1=-p_1$  & &   \\
$[e_2,e_4]=e_1 $ & $ Q_2=-x_2p_1$ &$H=Q_1=-p_1$  &$H,Q_2,Q_3$   \\
 $ [e_3,e_4]=e_2$ & $ Q_3=-\frac{x_2^2}{2} p_1 $ & & \\
 & $ Q_4=p_2 $ &  &   \\

\hline
 $A_{4,2}^{-1}  $ & $ Q_1=-p_1$  & $H=\frac{1}{Q_1 Q_2}=\frac{1}{x_2 p_1^2}$ &  \\
$[e_1,e_4]=-e_1 $ & $ Q_2=-x_2 p_1$ & or  &   $H,Q_1,Q_2,Q_3$ \\
 $ [e_2,e_4]=e_2$ & $ Q_3=-\frac{x_2}{2}  (Ln|x_2|) p_1 $ &  & \\
  $ [e_3,e_4]=e_2+ e_3$& $ Q_4=x_1 p_1+2x_2p_2 $ & $H=Q_2 exp({- \frac{Q_3}{Q_2}})=- x_2^{\frac{1}{2}} p_1$ &   \\
 \hline
   $A_{4,3} $ & $ Q_1=-p_1 $  &  &  \\
$[e_1,e_4]= e_1 $ & $ Q_2=-x_2 p_2 $ &  &   $H,Q_1,Q_2,Q_3$ \\
$ [e_3,e_4]=e_2$ & $ Q_3={x_2}  (Ln|x_2|) p_1 $ & $H=Q_1 exp({- \frac{Q_3}{Q_2}})=- x_2 p_1$ & \\
   & $ Q_4= - x_1 p_1- x_2 p_2 $ &  &   \\
\hline
     $A_{4,4} $ & $ Q_1=-p_1 $  &  &  \\
$[e_1,e_4]= e_1 $ & $ Q_2=-x_2 p_1 $ & $H=Q_1 exp({- \frac{Q_2}{Q_1}})$  &   $H,Q_1,Q_2,Q_3$ \\
 $ [e_2,e_4]=e_1+ e_2$ & $ Q_3=\frac{- 1}{2} x_2^2 p_1 $ &  & \\
  $ [e_3,e_4]=e_2+ e_3$ & $ Q_4= - x_1 p_1+ p_2 $ &  $\hspace{.4cm}=-exp({ x_2}) p_1$&   \\
\hline
     $A_{4,5} ^{a ,b ,1}$ &   &  &  \\
   & $ Q_1=-p_1 $ & &\\
$[e_1,e_4]= a e_1 $ & $ Q_2=- e^{(a-b) x2} p_1 $ & $H=\frac{Q_1^b}{Q_2}=\frac{p_1^{(b-1)}}{e^{(b-a) x_2}}$  &   $H,Q_1,Q_2,Q_3$ \\
 $ [e_2,e_4]= b e_2 $ & $ Q_3=- e^{(a-1) x2} p_1$ & or  & \\
  $ [e_3,e_4]=  e_3$ & $ Q_4= - a x_1 p_1- p_2 $ & $H=\frac{Q_1^b}{Q_2}=\frac{p_1^{(b-1)}}{e^{(a-1) x_2}}$ &   \\
   & & &\\
 $ -1\leq a < b < 1$ & & &\\
$ b > 0  ~~ if~ a=-1$  & & &\\

\hline  $A_{4,6}^{a,b}  $ & $ Q_1=-p_1$  & $H=\frac{Q_1^{ \frac{2b}{a}}}{Q_2^2+Q_3^2}$ &  \\
$[e_1,e_4]=a e_1 $ & $ Q_2=- e^{(a-b) x2} cos(x_2)  p_1$ & &   $H,Q_1,Q_2,Q_3$ \\
 $ [e_2,e_4]=b e_2 - e_3 $ & $ Q_3= e^{(a-b) x2} sin(x_2) p_1 $ &$=\frac{- p_1^{ \frac{2b}{a}}-2}{2 e^{2(a-b) x2} p_1 }$ & \\
  $ [e_3,e_4]=e_2+ b e_3$& $ Q_4=- a x_1 p_1- p_2 $ &  &   \\

$  b\geq 0 $& & &\\
$a \neq 0 $ & & &\\

\hline  $A_{4,7}  $ & $ Q_1=-p_1$  &  &  \\
$[e_1,e_4]=2 e_1 $ & $ Q_2=- x_2 p_1$ & &   $H,Q_1$ \\
 $ [e_2,e_4]= e_2 $ & $ Q_3= p_2 $ &$H=Q_2=- x_2 p_1$ & \\
  $ [e_3,e_4]=e_2+ e_3$& $ Q_4=-(2 x_1-1/2 x_2^2) p_1-x_2 p_2 $ &  &   \\
$ [e_2,e_3]= e_1 $& & &\\

\hline  $A_{4,9}^{b}  $  & & &\\
$[e_2,e_3]=e_1 $ & $ Q_1=-p_1$  &  &  \\
$[e_1,e_4]= (1+b)e_1 $ & $ Q_2=- p_2$ & $H=Q_1= - p_1 $ &   $H,Q_2$ \\
 $ [e_2,e_4]=e_2$ & $ Q_3=-x_2 p_1 $ &  & \\
  $ [e_3,e_4]=b e_3 $ & $ Q_4=-(1+b) x_1 p_1- x_2 p_2 $ &  &   \\
 $\mid  b\mid \leq 1$& & &\\

 \hline  $A_{4,12}$  & & &\\
$[e_1,e_3]=e_1 $ & $ Q_1=-p_1$  & & \\
$[e_2,e_3]=e_2 $ & $ Q_2=- x_2 p_1$ &  $H=Q_2=- x_2 p_1$  &   $H,Q_1$ \\
 $ [e_1,e_4]=-e_2$ & $ Q_3=- x_1 p_1 $ &    &  \\
  $ [e_2,e_4]=e_1$& $ Q_4= x_1 x_2 p_1+(1+x_2^2) p_2 $ &  &   \\
\hline
\end{tabular} }

\newpage
{
\qquad Table 2:{\tiny{Integrable and superintegrable systems with the  phase space $\mathbb {R}^{6}$ .}}\\
{\tiny{}
\begin{tabular}{|p{2.78cm}|p{0.07cm}|p{4.5cm}|p{3.55cm}|p{1.65cm}|}
\hline {symmetry Lie algebra} & & & &\\
{ (nonzero commutation}  & N &\hspace{2cm} $Q_i$&\hspace{1.55cm}$ H$& invariants  \\
\qquad relations) & & & &\\\hline

 \hline $A_{4,1}  $& 1 & $ Q_1=-p_1 $ &   &    \\
$[e_2,e_4]=e_1 $& & $ Q_2=-p_2$ & $ H=Q_2^2-2Q_1Q_3$ & $ H,Q_1,Q_2,Q_3$  \\
 $ [e_3,e_4]=e_2$& & $ Q_3=-p_3 $ &  & \\
 & & $ Q_4=- x_2 p_1 -x_3 p_2 $ &$\hspace{0.4cm}=p_2^2-p_1p_3$ &  \\
\cline{2-5}
&2  & $ Q_1=-p_1 $ & &    \\
 & &$ Q_2=-p_2$ & $ H=Q_2^2-2Q_1Q_3 $  &  $ H,Q_1,Q_2,Q_3$ \\
  & &$Q_3=\frac{1}{2}x_3^2p_1-x_3p_2 $ &  & \\
  & &$ Q_4=- x_2 p_1 + p_3 $ &$\hspace{0.4cm}=p_2^2+\frac{1}{2}x_3^2p_1^2-x_3p_1p_2$ &  \\
\cline{2-5}
 &3 & $ Q_1=-p_1 $ &  &  \\
 && $ Q_2=-x_2p_1$ & $ H=Q_2^2-2Q_1Q_3$ &  $ H,Q_1,Q_2,Q_3 $   \\
  && $Q_3=-p_3 $ & & \\
  & & $ Q_4=- x_2x_3 p_1 + p_2 $ &$\hspace{0.4cm}=x_2^2p_1^2-2 p_1p_3$  &  \\
\cline{2-5}
&4  & $ Q_1=-p_1 $ & &   \\
 && $ Q_2=-x_2p_1$ & $ H=Q_2^2-2Q_1Q_3$  & $ H,Q_1,Q_2,Q_3 $   \\
 & & $Q_3=-x_3p_1 $ &  & \\
 & & $ Q_4=p_2 + x_2p_3 $ &$\hspace{0.4cm}=(x_2^2-2 x_3)p_1^2$ &  \\
  \hline

 \hline $A_{4,2}^b  $&1 & $ Q_1=-p_1 $ &   &    \\
$[e_1,e_4]= b e_1 $ && $ Q_2=-p_2$ & $ H=Q_2 exp(- \frac{Q_3}{Q_2})$ & $ H,Q_1,Q_2,Q_3$  \\
 $ [e_2,e_4]=e_2$& & $ Q_3=-p_3$ && \\
 $ [e_3,e_4]=e_2+e_3 $& & $ Q_4=- b x_1 p_1- (x_2+x_3)p_2- x_3 p_3 $ & $\hspace{0.4cm}=-p_2 exp{(-\frac{p_3}{p_2})}$  &  \\
\cline{2-5}
 &2 & $ Q_1=-p_1 $ & &    \\
& & $ Q_2=-p_2$ & $ H=Q_2 exp(- \frac{Q_3}{Q_2})$   &  $ H,Q_1,Q_2,Q_3$ \\
 & & $Q_3=-x_3p_2 $ & $\hspace{0.4cm} =-p_2 exp{(-x_3)}$ & \\
 & & $ Q_4=- b x_1 p_1 - x_2 p_2 + p_3 $ & &  \\
\cline{2-5}
&3  & $ Q_1=-p_1 $ &  &  \\
 && $ Q_2=-x_2p_1$ &$ H=Q_2 exp(- \frac{Q_3}{Q_2})$  &  $ H,Q_1,Q_2,Q_3 $   \\
&  & $Q_3=- x_3p_1 $ & $\hspace{0.4cm}=- x_2p_1 exp{(-\frac{x_3}{x_2})}$ & \\
 & & $ Q_4=-b x_1 p_1-(b-1)x_2 p_2  $ & &  \\
&& $- ((b-1)x_3-x_2)p_3 $& &\\
\hline

 \hline $A_{4,3}  $&1 & $ Q_1=-p_1 $ &   &    \\
$[e_1,e_4]=  e_1 $& & $ Q_2=-p_2$ & $ H=Q_1 exp(- \frac{Q_3}{Q_2})$ & $ H,Q_1,Q_2,Q_3$  \\
& & $ Q_3=- p_3$ &  & \\
$[e_3,e_4]=e_2$ && $ Q_4=-  x_1 p_1- x_3 p_2 $ & $\hspace{0.4cm}=-p_1 exp{(-\frac{p_3}{p_2})}$ &  \\
\cline{2-5}
& 2 & $ Q_1=-p_1 $ & &    \\
& & $ Q_2=-x_2 p_1$ & $ H=Q_1 exp(- \frac{Q_3}{Q_2})$   &  $ H,Q_1,Q_2,Q_3$ \\
 & & $Q_3=-p_3 $ &  & \\
&  & $ Q_4=- (x_1+x_2 x_3) p_1 - x_2 p_2 $ & $\hspace{0.4cm}=-p_1 exp{(-\frac{p_3}{x_2 p_1})}$ &  \\
\cline{2-5}
& 3  & $ Q_1=-p_1 $ &  &  \\
& & $ Q_2=-x_2p_1$ & $ H=Q_1 exp(- \frac{Q_3}{Q_2})$ &  $ H,Q_1,Q_2,Q_3 $   \\
&  & $Q_3=- x_3p_1 $ && \\
&  & $ Q_4=-x_1 p_1-x_2p_2 - (x_3-x_2)p_3$ &  $\hspace{0.4cm} =- p_1 exp{(-\frac{x_3}{x_2})}$  &  \\
\hline

 \hline $A_{4,4}  $&1 & $ Q_1=-p_1 $ & $H=Q_1 exp(- \frac{Q_2}{Q_1})$  &    \\
$[e_1,e_4]=  e_1 $ & & $ Q_2=-p_2$ & $ \hspace{0.4cm} =-p_1 exp{(-\frac{p_2}{p_1})}$ & $ H,Q_1,Q_2,Q_3$  \\
$[e_2,e_4]=e_1+e_2$ && $ Q_3=- p_3$ & or & \\
$[e_3,e_4]=e_2 + e_3$& & $ Q_4=-(x_1+x_2)p_1-$ & $H=\frac{2Q_1Q_3-Q_2^2}{Q_1^2} $ &  \\
&& $(x_2+x_3)p_2-x_3p_3$&$\hspace{0.4cm}=\frac{2p_1 p_3-p_2^2}{p_1^2}$&\\
\cline{2-5}
& 2 & $ Q_1=-p_1 $ &$H=Q_1 exp(- \frac{Q_2}{Q_1})$ &    \\
& & $ Q_2=- p_2$ & $ \hspace{0.4cm}=-p_1 exp{(-\frac{p_2}{ p_1})}$   &  $ H,Q_1,Q_2,Q_3$ \\
&  & $Q_3=1/2 x_3^2 p_1 -x_3 p_2 $ & or & \\
&  & $ Q_4=-(x_1+x_2)p_1-x_2p_2+p_3$ &  $H=\frac{2Q_1Q_3-Q_2^2}{Q_1^2} $ &  \\
&&&$ \hspace{.4cm}=\frac{-x_3^2 p_1^2+2x_3p_1 p_2-p_2^2}{p_1^2}$&\\

\cline{2-5}
& 3 & $ Q_1=-p_1 $ & $H=Q_1 exp(- \frac{Q_2}{Q_1})$ &  \\
& & $ Q_2=-x_2p_1$ & $\hspace{0.4cm} =- p_1 exp{(-\frac{x_3}{x_2})}$ &  $ H,Q_1,Q_2,Q_3 $   \\
&  & $Q_3=- x_3p_1 $ & or & \\
&  & $ Q_4=-x_1 p_1+p_2 +x_2p_3$ &  $H=\frac{2Q_1Q_3-Q_2^2}{Q_1^2}=2x_3-x_2^2 $ &  \\
\hline

\end{tabular} }
\newpage
{
\qquad Table 2:{\tiny{Integrable and superintegrable systems with the  phase space $\mathbb {R}^{6}$ (continue).}}\\
{\tiny{}
\begin{tabular}{|p{2.54cm}|p{0.05cm}|p{4.5cm}|p{3.55cm}|p{1.65cm}|}

\hline $A_{4,5}^{a,b,c},~ a b c\neq 0 $&1 & $ Q_1=-p_1 $ &   &    \\
$[e_1,e_4]= a e_1 $& & $ Q_2=-p_2$ & $ H= \frac{Q_1^b}{Q_2}=\frac{(-p_1)^b}{p_2}$ & $ H,Q_1,Q_2,Q_3$  \\
$[e_2,e_4]= b e_2$& & $ Q_3=- p_3$ & or & \\
$[e_3,e_4]= c e_3$& & $ Q_4=-a x_1p_1- b x_2p_2-c x_3p_3$ &  $ H= \frac{Q_1^c}{Q_3}=\frac{(-p_1)^c}{p_3}$ &  \\
\cline{2-5}
&2  & $ Q_1=-p_1 $ & &    \\
& & $ Q_2=-x_2 p_1$ & $ H=- \frac{Q_1^b}{Q_2}=\frac{(-p_1)^{b-1}}{x_2}$  &  $ H,Q_1,Q_2,Q_3$ \\
 & & $Q_3=- x_3 p_1  $ & or & \\
&  & $ Q_4=- a x_1p_1-(a-b)x_2p_2-$ & $ H= \frac{Q_1^c}{Q_3}=\frac{(-p_1)^{c-1}}{x_3} $ &  \\
&&$(a-c) x_3 p_3$&&\\

\dots\dots\dots\dots\dots\dots\dots       &\dots &\dots\dots \dots \dots \dots \dots \dots \dots \dots\dots\dots\dots     &\dots\dots \dots \dots \dots \dots \dots \dots \dots \dots   & \dots\dots \dots \dots \\

$a=b=1$ &3 & $ Q_1=-p_1 $ &  &  \\
$c\neq1$ && $ Q_2=-x_2p_1$ & $ H= \frac{Q_1}{Q_2}=\frac{1}{x_2}$ &  $ H,Q_1,Q_2,Q_3 $   \\
&  & $Q_3=- p_3 $ & or & \\
&  & $ Q_4=-x_1 p_1 - c x_3p_3 $ & $ H= \frac{Q_1^c}{Q_3}=\frac{(-p_1)^c}{- p_3}$ &  \\
\dots\dots\dots\dots\dots\dots\dots       &\dots &\dots\dots \dots \dots \dots \dots \dots \dots \dots\dots\dots\dots     &\dots\dots \dots \dots \dots \dots \dots \dots \dots \dots   & \dots\dots \dots \dots \\

$-1\leq a< b<1$&4  & $ Q_1=-p_1 $ &  &  \\
$c=1$ && $ Q_2=-x_2p_1$ & $ H= \frac{Q_1^b}{Q_2}=\frac{(-p_1)^{b-1}}{x_2}$ &  $ H,Q_1,Q_2,Q_3 $   \\
$b>0 ~if~a=-1$ & & $Q_3=- p_3 $ & or & \\
& & $ Q_4=- a x_1p_1-(a-b)x_2p_2-x_3 p_3 $ &   $ H= \frac{Q_1}{Q_3}=\frac{p_1}{ p_3} $ &  \\
 \hline

\hline $A_{4,6}^{a,b} ,~a > 0 $ &1& $ Q_1=-p_1 $ &   &    \\
$[e_1,e_4]= a e_1 $ && $ Q_2=-p_2$ & $ H= \frac{Q_1^{\frac{2b}{a}}}{Q_2^2+Q_3^2}=\frac{(-p_1)^{\frac{2b}{a}}}{p_2^2+p_3^2}$ & $ H,Q_1,Q_2,Q_3$  \\
$[e_2,e_4]= b e_2 - e_3$& & $ Q_3=- p_3$ &  & \\
$[e_3,e_4]= e_2+b e_3$ & &$ Q_4=-a x_1p_1-( b x_2+x_3)p_2$ & &  \\
&&$-(-x_2+b x_3)p_3$&&\\
\cline{2-5}
&2  & $ Q_1=-p_1 $ & &    \\
 && $ Q_2=-x_2 p_1$ & $ H= \frac{Q_1^{\frac{2b}{a}}}{Q_2^2+Q_3^2}$  &  $ H,Q_1,Q_2,Q_3$ \\
&  & $Q_3=- x_3 p_1  $ &  & \\
&  & $ Q_4=- a x_1p_1-((a-b)x_2+x_3)p_2$ &$\hspace*{.4cm}=\frac{(-p_1)^{\frac{2(b-a)}{a}}}{x_2^2+x_3^2}$  &  \\
&  &$-(-x_2+(a-c) x_3) p_3$&&\\
\hline

 \hline $A_{4,9}^b ~,~~ |b|\leq 1 $ &1& $ Q_1=-p_1 $ &   &    \\
$[e_1,e_4]= (1+b) e_1 $& & $ Q_2=-p_2$ &$H=Q_1=p_1$  & $ H,Q_2$  \\
$[e_2,e_4]=e_2$ && $ Q_3=-x_2p_1- p_3$ & & \\
$[e_3,e_4]=b e_3$ && $ Q_4=-(1+b)x_1p_1-x_2p_2-b x_3p_3$ & &  \\
\cline{2-5}
 $[e_2,e_3]=e_1$ &2& $ Q_1=-p_1 $ & &    \\
& & $ Q_2=- p_2$ &  $H=Q_1=p_1$ &  $ H,Q_2$ \\
 & & $Q_3=- x_2 p_1 -x_3 p_2 $ & & \\
& & $ Q_4=-(1+b)x_1p_1-x_2p_2$ & &  \\
& &$-(1-b) x_3p_3$ & &\\
\cline{2-5}
&3  & $ Q_1=-p_1 $ &  &  \\
& & $ Q_2=-p_2$ &  $H=Q_1=p_1$&  $ H,Q_2 $   \\
 & & $Q_3=- x_2p_1 $ &  & \\
&  & $ Q_4=-(1+b)x_1p_1-x_2p_2- p_3$ &   &  \\
\hline

\hline $A_{4,12} $ &1& $ Q_1=-p_1 $ &   &    \\
$[e_1,e_3]=  e_1 $ && $ Q_2=-p_2$ & $H=Q_1=p_1$ & $ H,Q_2$  \\
$[e_2,e_3]=e_2$ && $ Q_3=-x_1p_1-x_2 p_2- p_3$ & & \\
$[e_1,e_4]=- e_2$& & $ Q_4=-x_2p_1+x_1p_2-C p_3$ & &  \\
\cline{2-5}
 $[e_2,e_4]=e_1$&2& $ Q_1=-p_1 $ & &    \\
& & $ Q_2=-x_2 p_1$ & $H=Q_1=p_1$ &  $ H,Q_2$ \\
&  & $Q_3=- x_1 p_1 - p_3 $ & & \\
 & & $ Q_4=x_1x_2p_1+(1+x_2^2)p_2$ &  &  \\
\cline{2-5}
 &3 & $ Q_1=-p_1 $ &  &  \\
& & $ Q_2=-p_2$ & $H=Q_1=p_1$&  $ H,Q_2 $   \\
  && $Q_3=- x_1p_1-x_2p_2 $ &  & \\
&  & $ Q_4=-x_2p_1+x_1p_2- p_3$ &   &  \\
\hline
\end{tabular}}\\
\section{Integrable and superintegrable Hamiltonian systems with the symmetry Lie group as  phase space of the system }

In this section, we construct the integrable Hamiltonian systems with the symmetry Lie group as a four dimensional phase space. For this propose, we consider those four dimensional real Lie groups such that they have symplectic structure. The list of symplectic four dimensional real Lie groups are classified in \cite{gp}. Here, we construct the models on those Lie groups separately as follows.\\
\textbf{ Lie group  $ \mathbf A_{4,1}$:}\\

According to \cite{gp}, \cite{ms} and \cite{bm}, non-degenerate
Poisson $P^{\mu  \nu}$ structure on this Lie group can be obtained
in  the following forms:\footnote{ Not that in \cite{gp} and
\cite{ms} the symplectic structure $\omega _{i j}$ on Lie algebra
have been given. For obtaining  the symplectic  structure
$\omega_{\mu \nu}=e_\mu ^{~i } \omega_{i j} e_\nu ^{~j}$on groups
one can use the vierbein $e_\mu ^{~i }$ which have been obtained in
\cite{bm} for four dimensional real Lie groups. Then, one can obtain
the non-degenerate Poisson structure from  $ P ^{\mu  \nu}=
(\omega_{\mu \nu})^t$}
 \begin{equation}\label{B1}
  \lbrace x_1,x_2\rbrace =-\frac{c}{2} x_4^2 ,~  \lbrace x_1,x_3\rbrace =c x_4 ,~  \lbrace x_1,x_4\rbrace =-d ,~  \lbrace x_2,x_3\rbrace =-c,
\end{equation}
where $ c $ and $ d $ are arbitrary real constants.\\
Now, one  can find the following Darboux coordinates:\\
$\hspace*{.65cm}y_1 =\frac{ x_3}{c} +\frac{ (c x_4^2)}{8 }+\frac{ x_4^2}{(2 d)}, \\
\hspace*{.65cm}y_2=-x_1 +\frac{ x_3^2}{c^2} + \frac{1}{4} c d x_2 x_4 -\frac {x_3 x_4^2}{4} + \frac{x_3 x_4^2}{c d} - \frac{3 c^2 x_4^4}{64} +\frac{ x_4^4}{4 d^2} -\frac {c x_4^4}{8 d} ,\\
\hspace*{.65cm}y_3 = x_2-\frac{2 x_3 x_4}{c d}-\frac{x_4^3}{d^2}-\frac{c x_4^3}{4 d} ,$\\
\begin{equation}\label{B2}
y_4=\frac{1}{d} x_4,\hspace*{13.5cm}
\end{equation}
such that  they satisfy  the following standard Poisson brackets:
\begin{equation}
 \lbrace y_1,y_3\rbrace =1~,~~~~ \lbrace y_2,y_4\rbrace =1 .
\end{equation}
In other words the coordinate $y_i$ can be used as a coordinates for the phase space $\mathbb {R}^{4}$; such that the $y_1$ and $y_2$ are dynamical variables and $p_{y_1}=y_3$ and $p_{y_2}=y_4$ are their momentum conjugate.  On the other hand, we can apply the realization  of $ A_{4,1}$ of table 1 with phase space $\mathbb {R}^{4}$ with coordinates $y_i$; in this respect, using (\ref{B2}) and after replacing in that realization  $y_i$ in terms of $x_i$   we obtain the following realization for $Q_i$:\\
$\hspace*{.65cm}Q_1=-x_2+\frac{2 x_3 x_4}{c d}+\frac{x_4^3}{d^2}+\frac{c x_4^3}{4d},\\
\hspace*{.65cm} Q_2= (x_1 -\frac{ x_3^2}{c^2} - \frac{1}{4} c d x_2
x_4 +\frac {x_3 x_4^2}{4} - \frac{x_3 x_4^2}{
   c d} + \frac{3 c^2 x_4^4}{64} -\frac{ x_4^4}{4 d^2} +\frac {c x_4^4}{8 d})\\ (x_2 - \frac{
   8 d x_3 x_4 + 4 c x_4^3 + c^2 d x_4^3}{4 c d^2}),\\
\hspace*{.65cm}Q_3=-\frac{1}{2} (-x_1 +\frac{ x_3^2}{c^2} +
\frac{1}{4} c d x_2 x_4 -\frac {x_3 x_4^2}{4} + \frac{x_3 x_4^2}{
   c d} - \frac{3 c^2 x_4^4}{64} +\frac{ x_4^4}{4 d^2} -\frac {c x_4^4}{8 d})^2\\ (x_2 - \frac{8 d x_3 x_4 + 4 c x_4^3 + c^2 d x_4^3}{4 c
   d^2}),$
\begin{equation}\label{Q1}
 Q_4=\frac{1}{d} x_4,\hspace*{13cm}
\end{equation}
such that they satisfy  the following Poisson brackets by  use of
(\ref{B1}) as
\begin{equation}
\{Q_2,Q_4\}=Q_1~,~~\qquad \{Q_3,Q_4\}=Q_2 .
\end{equation}
Then, the Hamiltonian of the \textit{superintegrable} system with
the $\mathbf A_{4,1}$ as a phase space and symmetry group is
obtained as follows:
\begin{equation}\label{Q2}
H=Q_1 = -x_2+\frac{2 x_3 x_4}{c d}+\frac{x_4^3}{d^2}+\frac{c
x_4^3}{4 d},
\end{equation}
where the invariants of the system are $(H,Q_2,Q_3)$.\footnote{Note that in the relation (\ref{Q1}) and (\ref{Q2}) and also the relations in the forthcoming models, one can choose the variables $x_1$ and $x_2$ as dynamical variables with momentum conjugates $p_{x_1}=x_3$ and $p_{x_2}=x_4$.}\\

\textbf{ Lie group $ \mathbf A_{4,2}^{-1}$:}\\

The non-degenerate Poisson structure on $ \mathbf A_{4,2}^{-1}$  can
be obtained as follows \cite{gp}, \cite{ms}, \cite{bm}:
\begin{equation}\label{D1}
 \{x_1,x_2\}=2 a ,\quad\{x_1,x_3\}=-a ,\quad\{x_2,x_4\}= b~ e^{-x_4}~,
 \end{equation}
 where $ a $ and $ b $ are arbitrary real constants. For this example, the Darboux coordinates has the following forms:
$$ y_1=-\frac{e^{x_4}}{b}+x_3   ,\hspace{1.75cm}\qquad y_2=\frac{-2 a e^{x_4 }- b x_1 + a b x_2}{a b^2},$$
  \begin{equation}
  y_3=\frac{2 e^{x_4}}{b} + \frac{x_1}{a}~~~~,~~~~~~~~\qquad y_4=e^{x_4 }.\hspace{3cm}
  \end{equation}
Then, after using the results of  table 1, we have the following forms for the dynamical functions $ Q_i $:\\
$\hspace*{.65cm}Q_1 =-\frac{2 e^{x_4}}{b} - \frac{x_1}{a},\\
\hspace*{.65cm}Q_2 =-( \frac{-2 a e^{x_4 }- b x_1 + a b x_2}{a b^2})(\frac{2 e^{x_4}}{b} + \frac{x_1}{a}),\\
\hspace*{.65cm}Q_3 =-\frac{1}{2}(\frac{-2 a e^{x_4 }- b x_1 + a b
x_2}{a b^2})(\frac{2 e^{x_4}}{b} + \frac{x_1}{a}) Ln(\vert \frac{-2 a
e^{x_4 }- b x_1 + a b x_2}{a b^2}\vert )$,
\begin{equation}
Q_4=2 e^{x_4 }(\frac{-2 a e^{x_4 }- b x_1 + a b x_2}{a b^2})
+(-\frac{e^{x_4}}{b}+x_3 )(\frac{2 e^{x_4}}{b} +
\frac{x_1}{a}),\hspace{4.3cm}
\end{equation}
such that they satisfy  the following Poisson brackets by  use of
(\ref{D1}) as
\begin{equation}
\qquad \{Q_1,Q_4\}=-Q_1~,~~~ \{Q_2,Q_4\}=Q_2~,~~~
\{Q_3,Q_4\}=Q_2+Q_3 ,
\end{equation}
In this respect, the Hamiltonian of the \textit{maximal
superintegrable} system with the $\mathbf A_{4,2}^{-1}$ as a phase
space and symmetry group is obtained as follows:
\begin{equation}
H=\frac{1}{Q_1 Q_2}= \frac{1} {(\frac{2 e^{x_4}}{b} +
\frac{x_1}{a})^2(\frac{-2 a e^{x_4 }- b x_1 + a b x_2}{a b^2})} ~,
\end{equation}
where the invariants of the system are $(H,Q_1,Q_2,Q_3)$.\\

\textbf{ Lie group $\mathbf A_{4,3}$:}\\

From \cite{gp}, \cite{ms} and \cite{bm}, we have the following forms
for the non-degenerate Poisson structure on $\mathbf A_{4,3}$:
$$\{x_1,x_2\}= c~ x_4 e^{-x_4}~ ,\hspace{1cm}\{x_1,x_3\}=d ~e^{-x_4 },$$
\begin{equation}\label{E1}
 ~\{x_1,x_4\}=h e^{-x_4 }~~,\hspace{1.5cm}\{x_2,x_3\}=f, \hspace{1cm}
 \end{equation}
 where $c, d, h $ and $ f $ are arbitrary real constants.\\
Now, after finding of Darboux coordinates in the following forms:\\
$\hspace*{1.5cm} y_1=\frac{d x_2}{f} + \frac{c h x_3^2}{2 d f} -
\frac{c x_3
x4}{f},\quad\qquad y_2=\frac{x_1}{h} - \frac{d e^{-x_4} x_2}{f h} - \frac{c e^{-x_4} x_3^2}{2 d f} + \frac{ c e^{-x_4} x_3 x_4}{f h},$\\
\begin{equation}
 y_3=\frac{x_3}{d}, \hspace*{4cm} y_4=e^{x_4}.\hspace*{5.75cm}\\
\end{equation}
one can obtain the $ Q_i $  as follows:\\
$\hspace*{.65cm}Q_1=-\frac{x_3}{d}, \\
\hspace*{.65cm}Q_2=\frac{x_3}{d}(-\frac{x_1}{h} + \frac{d e^{-x_4 }x_2}{f h} + \frac{c e^{-x_4} x_3^2}{2 d f} - \frac{ c e^{-x_4}x_3 x_4}{f h}),\\
\hspace*{.65cm}Q_3=\frac{x_3}{d}(\frac{x_1}{h} - \frac{d e^{-x_4}
x_2}{f h} - \frac{c e^{-x_4} x_3^2}{2 d f} + \frac{ c e^{-x_4} x_3
x_4}{f h})(Ln(\vert\frac{x_1}{h} - \frac{d e^{-x_4} x_2}{f h} -
\frac{c e^{-x_4} x_3^2}{2 d f} + \frac{ c e^{-x_4} x_3 x_4}{f
h}\vert)),$
\begin{equation}
Q_4=-e^{x_4} \frac{x_1}{h} +\frac{(d - h x_3) (2 d^2 x_2 + c h x_3^2
- 2 c d x_3 x_4)}{2 d^2hf},\hspace*{6.2cm}
\end{equation}
such that they satisfy  the following Poisson brackets by  use of
(\ref{E1}) as
\begin{equation}
 \{Q_1,Q_4\}=Q_1~,~~\qquad \{Q_3,Q_4\}=Q_2.
 \end{equation}
Then the Hamiltonian of the\textit{ maximal superintegrable } system
with the $\mathbf A_{4,3}$ as a phase space and symmetry group is
obtained as
\begin{equation}
H=Q_1 exp(-\frac{Q_3}{Q_2})= \frac{e^{-x_4} x_3}{2 d^2 f h}(c h
x_3^2 - 2 d (f e^{x_4}  x_1 - d x_2 + c x_3 x_4)),
\end{equation}
where the invariants of the system are $(H,Q_1,Q_2,Q_3)$.\\

\textbf{ Lie group $\mathbf A_{4,6}^{a,0}$:}\\

For this Lie group we have the following    non-degenerate Poisson  structure \cite{gp}, \cite{ms}, \cite{bm}: \\
\begin{equation}\label{F1}
 \quad\{x_1,x_4\}=d~ e^{-a x_4},\quad\{x_2,x_3\}=c\quad ,
 \end{equation}
 where $ c $ and $ d $ are arbitrary real constants.The Darboux coordinates for this structure are as follows:
 \begin{equation}
 y_1=x_3~,\qquad
 y_2=-\frac{e^{2 a x4} x_1}{a d}~,\qquad
 y_3=-\frac{x_2}{c}~,\qquad
 y_4=e^{-a x_4},
 \end{equation}
 such that after the same calculation and using the results  of table 1, the $ Q_i $ have the following forms:\\
$\hspace*{.65cm}Q_1=\frac{x_2}{c}, \\
\hspace*{.65cm}Q_2=\frac{e^{-(\frac{e^{2 a x_4} x_1}{d})}~ x_2 ~cos(\frac{e^{2 a x_4} x_1}{a d})}{c},\\
\hspace*{.65cm}Q_3=\frac{e^{-(\frac{e^{2 a x_4} x_1}{d})} ~x_2~
sin(\frac{e^{2 a x_4} x_1}{a d})}{c},$
\begin{equation}
Q_4=-e^{-a x_4} + \frac{a}{c} x_2 x_3 .\hspace*{11cm}
\end{equation}
where they satisfy the following Poisson brackets by  use of
(\ref{F1})
\begin{equation}
 \{Q_1,Q_4\}= a ~Q_1~,~~\qquad \{Q_2,Q_4\}=-Q_3~~\qquad \{Q_3,Q_4\}=Q_2.
\end{equation}
 The Hamiltonian of the \textit{maximal superintegrable} system with the $\mathbf A_{4,6}^{a,0}$ as a phase space and symmetry group
 is obtained
\begin{equation}
H=Q_2^2 + Q_3^2=\frac{e^{-(\frac{2e^{2 a x_4} x_1}{d})} x_2^2 }{c^2},
\end{equation}
where the invariants of the system are $(H,Q_1,Q_2,Q_3)$.\\

 \textbf{ Lie group $\mathbf A_{4,7}$:}\\

 The  non-degenerate Poisson structure for this Lie group has the following form \cite{gp}, \cite{ms}, \cite{bm}:
 \begin{equation}\label{G1}
\{x_1,x_3\}=- 2c x_3 e^{-2 x_4}~,~\{x_1,x_4\}=c e^{-2
x_4}~,~\{x_2,x_3\}=2c e^{-2 x_4}~,
 \end{equation}
 where $ c $ is the  arbitrary real constant. Furthermore, for this example one can find the following Darboux coordinates:\\
 $$ y_1= \frac{e^{2 x_4} ( x_2 )}{2 c}~,\hspace*{1.2cm} y_2=- \frac{ -1 - e^{2 x_4} + e^{4 x_4} x_1 + e^{4 x_4} x_2 x_3}{2 c},$$
\begin{equation}
y_3=x_3~~~,\hspace{2cm} y_4=e^{-2 x_4},\hspace{4.75cm}
\end{equation}
such that the $ Q_i $ have the following forms:\\
$\hspace*{.65cm}Q_1=-x_3 ,\\
\hspace*{.65cm}Q_2=\frac{x_3 (-1 - e^{2 x_4} + e^{4 x_4} x_1 + e^{4
x_4} x_2 x_3)}{2 c}, $
\begin{equation}
Q_3=e^{-2 x_4}, \hspace*{13cm}
\end{equation}
$\hspace*{.65cm}Q_4=x_3( -\frac{e^{2 x_4} ( x_2 )}{ c}+\frac{ (-1 - e^{2 x_4} + e^{4 x_4} x_1 + e^{4 x_4} x_2 x_3)^2}{8 c^2})+\frac{e^{-2 x_4} (-1 - e^{2 x_4} + e^{4 x_4} x_1 + e^{4 x_4} x_2 x_3)}{2 c}, $\\

so that they satisfy the following Poisson brackets by  use of (\ref{G1}) as\\
$$ \{Q_2,Q_3\}= Q_1~,\quad \{Q_1,Q_4\}=2Q_1~,\quad \{Q_2,Q_4\}=Q_2,$$
\begin{equation} \{Q_3,Q_4\}=Q_2+Q_3.\end{equation}
Then, the Hamiltonian of the \textit{integrable} system with the $\mathbf A_{4,7}$ as a phase space and symmetry group is obtained\\
\begin{equation}
H=Q_2=\frac{x_3 (-1 - e^{2 x_4} + e^{4 x_4} x_1 + e^{4 x_4} x_2
x_3)}{2 c},
\end{equation}
where the invariants of the system are $(H,Q_1)$.\\

 \textbf{ Lie group $\mathbf A_{4,9}^1$:}\\

 For this Lie group the non-degenerate Poisson structure has the following form \cite{gp}, \cite{ms}, \cite{bm}:
 \begin{equation}\label{H1}
 \{x_1,x_3\}=2c x_3 e^{-2 x_4},~\{x_1,x_4\}=-c e^{-2 x_4},~\{x_2,x_3\}=-2c e^{-2 x_4},
 \end{equation}
 where $ c $ is arbitrary real constant. On the other hand, after the same calculation one can find the Darboux coordinates as follows:\\
  $$ y_1=- \frac{e^{2 x_4} ( x_2 )}{2 c}~,\hspace*{1cm} y_2=\frac{ -1 - e^{2 x_4} + e^{4 x_4} x_1 + e^{4 x_4} x_2 x_3}{2 c},$$
 \begin{equation}
 y_3=x_3~~~,\hspace{2.2cm} y_4=e^{-2
x_4},\hspace{4.5cm}
  \end{equation}
Then, according to the results of table 1,  the $ Q_i $ are obtained as follows:\\
$\hspace*{.65cm}Q_1= - x_3,\\
\hspace*{.65cm}Q_2=-e^{-2 x_4},\\
\hspace*{.65cm}Q_3=-\frac{x_3 (-1 - e^{2 x_4} + e^{4 x_4} x_1 + e^{4 x_4} x_2 x_3)}{ 2 c}$ ,\\
\begin{equation}
Q_4= \frac{e^{2 x_4} ( x_2 x_3)}{ c}-\frac{e^{-2 x_4} (-1 - e^{2
x_4} + e^{4 x_4} x_1 + e^{4 x_4} x_2 x_3)}{ 2 c}, \hspace*{4.8cm}
\end{equation}

such that they satisfy  the  following Poisson brackets by  use of (\ref{H1}) as\\
$$ \{Q_2,Q_3\}= Q_1~,\quad \{Q_1,Q_4\}=2Q_1~,\quad \{Q_2,Q_4\}=Q_2 ,$$
\begin{equation} \{Q_3,Q_4\}=Q_3\end{equation}
In this way, the Hamiltonian of the\textit{ integrable}  system with
the $\mathbf A_{4,9}^1$ as a phase space and symmetry group is
obtained
\begin{equation}
H=Q_1=-x_3,
\end{equation}
where the invariants of the system are $(H,Q_2)$.\\

\textbf{ Lie group $\mathbf A_{4,12}$:}\\

Finally, for this Lie group we have the following  non-degenerate Poisson structure \cite{gp}, \cite{ms}, \cite{bm} :\\
$\hspace*{.65cm}\{x_1,x_3\}=-c ~ e^{- x_3}(a~cos(x_4)+b~sin(x_4)),\\
\hspace*{.65cm} \{x_1,x_4\}=c ~  e^{- x_3}(-b~cos(x_4)+a~sin(x_4)),\\
\hspace*{.65cm}\{x_2,x_3\}=c~ e^{- x_3}(b~ cos(x_4)-a~ sin(x_4)) ,$
\begin{equation}\label{I1}
 ~~ \{x_2,x_4\}=- c ~  e^{- x_3}(a~ cos(x_4)+b~ sin(x_4)),\hspace{7.65cm}
\end{equation}
where $c = \frac{1}{a^2+b^2}$ and  $a,b $  are arbitrary real constants   . One can find the following Darboux coordinates for this structure:\\
$\hspace*{.65cm}y_1=e^{2 x_3}(a x_1 cos(x_4) - b x_2 cos(x_4) +b x_1 sin(x_4) + a x_2 sin(x_4)) ,\\
\hspace*{.65cm} y_2=-e^{ x_3}(b x_1 cos(x_4) + a x_2 cos(x_4) - a x_1 sin(x_4) + b x_2 sin(x_4)) ,\\
 \hspace*{.65cm}y_3=e^{ x_3} ,$
\begin{equation}
 y_4=x_4.\hspace*{13.5cm}
 \end{equation}
Then, by use of table 1 one can obtain the $ Q_i $ as follows:\\
$\hspace*{.65cm}Q_1=-e^{- x_3},\\
\hspace*{.65cm}Q_2=b x_1 cos(x_4) + a x_2 cos(x_4) - a x_1 sin(x_4) + b x_2 sin(x_4),$\\
\begin{equation}
~~~~Q_3=-e^{ x_3}(a x_1 cos(x_4) - b x_2 cos(x_4) +b x_1 sin(x_4) +
a x_2 sin(x_4)),\hspace*{4.5cm}
\end{equation}
$\hspace*{.65cm}Q_4=- e^{2 x_3}(a x_1 cos(x_4) - b x_2 cos(x_4) +b x_1 sin(x_4) + a x_2 sin(x_4))(b x_1 cos(x_4) + a x_2 cos(x_4) - a x_1 sin(x_4) + b x_2 sin(x_4))+x_4(1-  e^{2 x_3} (b x_1 cos(x_4) + a x_2 cos(x_4) - a x_1 sin(x_4) + b x_2 sin(x_4))) $,\\
such that they satisfy  the following Poisson brackets by  use of (\ref{I1}) as\\
$$ \{Q_1,Q_3\}= Q_1~,\qquad \{Q_2,Q_3\}=Q_2~,\qquad \{Q_1,Q_4\}=-Q_2,$$
 \begin{equation}
 \{Q_2,Q_4\}=Q_1 .
 \end{equation}
Then, the Hamiltonian of the \textit{integrable} system with the
$\mathbf A_{4,12}$ as a phase space and symmetry group is obtained
\begin{equation}
H=Q_1=-e^{-2 x_4}
\end{equation}
where the invariants of the system are $(H,Q_2)$.

\bibliographystyle{amsplain}

\end{document}